# Eumelanin-based colouration reflects local survival of juvenile feral pigeons in an urban pigeon house.

Charlotte Récapet[1,2,3], Lise Dauphin[4], Lisa Jacquin[1,5], Julien Gasparini[1], Anne-Caroline Prévot-Julliard[4]

[1]*Lab. EcoEvo, UMR 7625 CNRS-UPMC-ENS, Univ. Pierre et Marie Curie, 7 quai St Bernard, 75252 Paris, France.*

[2]*LBBE, UMR 5558 CNRS-UCBL, Univ. Claude Bernard, Bât. Gregor Mendel, 43 boulevard du 11 novembre 1918, 69622 Villeurbanne, France*

[3]*Département d'Ecologie et d'Evolution, Biophore, Université de Lausanne, 1015 Lausanne-Dorigny, Switzerland*

[4]*Lab. CERSP, UMR 7204 CNRS-MNHN-UPMC, Muséum National d'Histoire Naturelle, CP 51, 55 rue Buffon, 75005 Paris, France*

[5]*Biology department & Redpath Museum, McGill University, 859 Sherbrooke Street W, Montréal, Québec*

**Corresponding author:**

Charlotte Récapet

Laboratoire de Biométrie et Biologie Evolutive,

Université Claude Bernard, CNRS, UMR 5558

Bât. Gregor Mendel, 43 boulevard du 11 novembre 1918, 69622 Villeurbanne, France

Tel: +33 (0)4 72 43 14 04

Fax: +33 (0)4 72 43 13 88

Email: charlotte.recapet@normalesup.org

**Abstract**

Urbanisation introduces deep changes in habitats, eventually creating new urban ecosystems where ecological functions are driven by human activities. The higher frequency of some phenotypes in urban vs. rural/wild areas has led to the assumption that directional selection in urban habitats occurs, which may thereby favour some behavioural and physiological traits in urban animal populations compared to rural ones. However, empirical evidence of directional selection on phenotypic traits in urban areas remains scarce. In this study we tested whether eumelanin-based colouration could be linked to survival in two urban populations of the feral pigeon, *Columba livia*. A number of studies in different cities pointed out a higher frequency of darker individuals in more urbanised areas compared to rural ones. To investigate whether directional selection through survival on this highly heritable trait could explain such patterns, we conducted mark-recapture studies on two populations of feral pigeons in highly urbanized areas. We predicted that darker coloured individuals would exhibit higher survival and/or philopatry (integrated into 'local survival') than paler coloured ones. No difference in local survival was found between adults of different colouration intensities. However, on one site, we found that darker juveniles had a higher local survival probability than light ones. Juvenile local survival on that site was also negatively correlated with the number of chicks born. This suggests the existence of colour- and/or density-dependent selection processes acting on juvenile feral pigeons in urban environments, acting through differential mortality and/or dispersal.

**Keywords**: *Columba livia*, feral pigeon, melanin-based colouration, juvenile survival, natal dispersal, colour polymorphism, urban ecosystems

# Introduction

The generalized urban sprawl in the 20th century is responsible for the replacement of rural or natural habitats by new urban ecosystems (Berry 1990), which are highly dependent on human-driven processes (Pickett et al. 1997, Grimm et al. 2000, Shochat et al. 2006). As a consequence, many species have declined in urban habitats, whereas others now successfully exploit these ecosystems (Kark et al. 2007). Consequently, species richness and assemblages are highly variable along urbanisation gradients (Clergeau et al. 1998, Germaine et al. 1998, Blair 2004). Moreover, animal communities exploiting urban habitats exhibit different characteristics compared to their rural counterparts, both at the species (Møller 2009) and at the population level (Luniak 2004, Evans 2009). For instance, urban populations usually have a longer reproductive season (Blair 2004, Partecke et al. 2004) and higher immune capacity (French et al. 2008, Fokidis et al. 2008), but lower anti-predatory defences (Blair 2004, Møller 2009) compared to rural populations of the same species (reviewed in Evans 2009). In addition, urban and rural populations show strong divergence in behavioural traits. For instance, individuals living in urban areas are more tolerant to human presence (Møller 2010), have higher innovation capacity or live in higher density populations compared to rural ones (Erz 1964, Luniak 2004, Møller 2009). Luniak et al. (2004) suggested that directional selection in urban areas could explain such phenotypic divergence on several physiological, morphological and behavioural traits between rural and urban individuals, resulting in a so-called "urbanization syndrome" and in local adaptation to urban areas (Evans 2009). However, empirical studies investigating the evolutionary processes underpinning phenotypic divergence between urban and rural animal populations are still scarce (Shochat et al. 2006, Evans 2009).

Melanin-based colouration is a good candidate trait to investigate urbanization syndrome, as it has a strong genetic basis (Roulin 2004), and a number of studies have reported differences in animal melanin-based colouration between urban and rural areas (Obukhova 2001, Yeh 2004). The degree of melanin-based colouration in vertebrates has been frequently reported to co-vary with other phenotypic traits such as immunity, behaviour or reproduction (reviewed in Ducrest et al. 2008), suggesting that melanin-based colouration reflects adaptations to different environmental conditions (Roulin 2004). For instance, a darker eumelanic colouration is often associated with higher aggressiveness (Reyer et al. 1998, Quesada and Senar 2007), a lower response to stress (Johnston and Janiga 1995, Almasi et al. 2010), a higher parasite resistance (Roulin et al. 2000; Jacquin et al. 2011), and a higher reproductive rate under food limitation compared to paler ones (Roulin et al. 2008, Jacquin et al. 2012). Such patterns have been suggested to be due to pleiotropic effects of genes coding for melanocortin ligands or receptors (Ducrest et al. 2008). As traits associated with a darker colouration match those found in individuals exploiting urban and/or stressful areas, melanin-based colouration might reflect an adaptation to the urban environment (Johnston and Janiga 1995, Jacquin 2011). As a consequence, one could expect directional selection to act on this trait and fitness of dark individuals should be higher than that of pale individuals in urban areas, a difference that could stem from differences in survival, reproductive success and habitat choice. Such differences in survival and/or dispersal according to colouration have already been described in birds (Roulin and Altwegg 2007, Roulin et al. 2010, Van den Brink et al. 2012, Saino et al. 2013).

In this study, we tested whether survival and/or dispersal differed according to melanin-based colouration in the urban feral pigeon *Columba livia*. Feral pigeons, originating from domesticated forms of the wild Rock Pigeon, have successfully colonized most cities of the world and have become one of the most common urban species (Haag-Wackernagel 1998, Kark et al. 2007). Interestingly, part of the colour variation selected for aesthetic purposes during the domestication process is maintained today in feral populations. This variation is strongly heritable (Jacquin et al. 2013a). Several explanations have been proposed such as differential adaptation to urban conditions (Johnston and Janiga 1995), parasites (Jacquin et al. 2011, Jacquin et al. 2013, Jacquin et al. in press) or food availability (Jacquin et al. 2012). However the link between colouration and fitness remains elusive (Leiss and Haag-Wackernagel 1999). A survey conducted at 192 European locations showed a positive correlation between the urbanization level and the proportion of dark eumelanic phenotypes (Obukhova 2007). The same correlation was found within the urban areas of Moscow (Obukhova 2011), Paris (Jacquin 2011, Jacquin et al. 2013b) and Manchester (S. L. Johnson, unpublished results). Such results suggest differences in fitness and/or habitat choice between dark and pale pigeons in urban areas. They are thus in agreement with our hypothesis that melanin-based colouration reflects an adaptation to the urban environment or to stressful habitats in general (Roulin et al. 2008).

Local survival probability, i.e., the probability to stay both alive and present in the sampled population, is a proxy of individual survival and/or dispersal. In this paper, we estimated local survival of feral pigeons roosting in two managed pigeon houses in urban areas near Paris, France, through a mark-recapture survey. We captured differently coloured pigeons at different ages (juveniles and adults) and estimated monthly local survival probabilities according to age, colour and number of chicks born in the pigeon houses. If a darker melanin-based colouration reflects adaptation to the urban environment, we predict that local survival would be higher in dark than in pale individuals.

## Material and methods

### Study sites

The study was conducted between 2007 and 2009 in two pigeon houses near Paris, in Pantin (Seine-Saint-Denis, France) and Fontenay (Val-de-Marne, France). The pigeon houses were built in 2003 and 1999, respectively, and had been freely colonized by feral pigeons within one year. Microsatellite data have shown that there is no genetic differentiation between pigeon flocks sampled within and around Paris (G. Jacob, A.C.Prévot-Julliard and E. Baudry, unpublished data). The sampled populations were about 7 km apart and should thus be genetically similar. These locations are both highly urbanized areas, measured as the proportion of built surfaces within 1 km of the pigeon houses (92.4% in Fontenay and 85.2% in Pantin; Jacquin et al. 2013b). In these public pigeon houses, pigeons were fed and eggs were removed in order to control reproduction (Jacquin et al. 2010). The two pigeon houses displayed the same density and size of available nest boxes and the same amount of food was provided in both pigeon houses (i.e. 50 kg of seeds once a week) during the study period. Feral pigeons lay 1 to 6 clutches of two eggs per year (Johnston and Janiga 1995). In Pantin, all newly laid eggs were removed in 2007, then one egg per clutch was removed in 2008 and 2009 (Table 1). In Fontenay, no eggs were removed in 2007 and 2008, then one egg per clutch was removed in 2009 (Table 1). These management procedures have been shown to cause changes in the reproduction phenology in these populations (Jacquin et al. 2010). In addition, they resulted in strong differences in the number of chicks born (Table 1) and juveniles fledged (Table 2). We thus included the number of chicks born in our analyses to take these differences into account.

*Capture and survey procedure*

Adult pigeons, which feed or sleep in the pigeon houses, were captured twice a year with a net tunnel placed in front of the entrance of the structure. Between 2007 and 2009, reproduction in both pigeon houses was monitored every week and juveniles born in the pigeon houses were ringed when 15-day old, about two weeks before leaving the nest. All individuals were equipped with a transponder included in one of the plastic rings. Individuals were detected by a receptor placed at the entrance of the pigeon house. Recording was conducted from August 2007 to December 2009 in Pantin (269 individuals surveyed) and from February 2007 to December 2009 in Fontenay (288 individuals surveyed). These recordings were then pooled to obtain monthly presence data. Due to the failure of receptors, no data were available for some months, at different times in the two sites (Table 1). We therefore analysed the data from the two sites separately. We let every individual's capture-recapture history start with the first transponder record of the individual. Therefore, juveniles were only surveyed from the first time that they left the pigeon house (i.e. when approximately one-month old), and survival during their first month was therefore not modelled.

*Colouration measurement*

Melanin-based colourations in feral pigeons are due to two types of pigment: pheomelanic pigments which are responsible for red colouration and eumelanic pigments, which are responsible for black and grey colouration (Haase et al. 1992). In this study we focused on eumelanin-based colouration, which is the most widespread colouration in feral pigeon populations, and thus excluded reddish and white individuals from the analyses (N = 22 and N = 6 respectively). Eumelanin deposition in feathers can vary in intensity, from total absence of pigmentation to fully dark melanic colouration (Haase et al. 1992). Although pheomelanin is present in "greyish" pigeons, no correlation with the intensity of eumelanin deposition could be found (Haase et al. 1992). We divided the continuous variation in four groups, following the classification used by Johnston and Janiga (1995) and Jacquin et al. (2011): (1) Blue bar (grey mantle with two black wing bars i.e. wild type); (2) Checker (a checked mantle with moderate black spots); (3) T-pattern (a black mantle with small grey marks); and (4) Spread (a completely black plumage). This classification reflects genetic differences at two multi-allelic loci (Johnston and Janiga 1995, Haag-Wackernagel et al. 2006), thus defining colour morphs (Roulin 2004). Although juveniles are slightly darker before their first moult, the morph of an individual remains the same throughout life (Johnston and Janiga 1995), an observation that was confirmed by photographs taken at different ages in captivity (L. Jacquin, pers. obs.). Previous studies have shown that eye scoring is a reliable method to estimate the colouration intensity of adult feral pigeons: it is highly repeatable and correlated with the percentage of dark feathers on the wing measured on standardized photographs (Jacquin et al. 2011).

*Local survival analyses*

To analyse recapture histories, we used Mark-Recapture statistical methodology to estimate two kinds of parameters: local survival probabilities $\Phi$ and recording probabilities $p$ (Lebreton et al. 1992). Recording probability is the probability for an individual alive and present to be recorded at a particular time point. Local survival probability is the probability to be still alive and present on the study site between two time points (thus referring to both survival and dispersal). In our study, these parameters were estimated on a monthly scale. Adult and juvenile recording histories within each site were analysed separately, as a function of age (in months for juveniles) and plumage colouration score. To get a sufficient sample size within each colouration score, we pooled morphs 3 and 4 in adults and morphs 2, 3 and 4 in juveniles (Table 2). Separating these groups did not improve the models indicating that there was either no difference or not enough power to detect a difference between these groups. It should also be noted that pooling morphs 2, 3 and 4 in adults did not change our results. Juvenile mortality occurs mostly before the age of six months (Johnston and Janiga

1995), and preliminary studies showed that juveniles older than five months have similar survival rate as adults (unpublished results). We thus tested six different age structures for each juvenile local survival model: age was described by a number of age-classes varying from one (no effect of age) to six (age effect up to six months). For example, five-month old chicks could either be considered as a distinct age class (six age-classes model) or be pooled with older individuals (one to five age-classes models).

We modelled the time-dependence in local survival by testing the effects of month, year and the interaction between the two. Since only one chick was born in 2007 in Pantin, it was not possible to estimate local survival rates and that year was excluded from the analysis for juveniles. Based on the results in Hetmański (2007), we included in our modelling the effect of juvenile density on local survival. We modelled the month by year effect on survival either as a categorical factor or as a numerical variable dependent on the number of chicks born during each month. In Fontenay where three years of data were available, we similarly modelled the year effect either as a categorical factor or as the total number of chicks born during each year.

The time-dependent model $\Phi_t\ p_t$ for each colour-group was first tested for goodness-of-fit (Pollock et al. 1985) in U-CARE 2.3.2 (Choquet et al. 2009a). A transience effect was detected in adults from both sites (Pantin: $z = 2.53$, $P = 0.01$; Fontenay: $z = 3.80$, $P = 0.001$), which means that local survival rate of newly marked individuals differed from that of previously marked ones. It could be explained either by the occurrence of 'transient' individuals that are only present during a short time in the study population, by an effect of marking on survival or dispersal, or by a heterogeneity in individual recording probabilities. We added one virtual age-class when modelling local survival probabilities to take this effect into account (Brownie and Robson 1983). Trap-dependence was significant for all site-by-stage categories except Pantin's juveniles (Fontenay: adults $z = -10.28$, $P < 0.001$; juveniles $z$

= -8.24, $P < 0.001$; Pantin: adults $z = -11.58$, $P < 0.001$; juveniles $z = -0.97$, $P = 0.33$). To take this effect into account, we modelled different recording probabilities according to the previous recording event for each individual in a multi-events model (Pradel 1993).

We therefore defined three states (alive and recorded, alive but not recorded, dead), and two events (recorded, not recorded). Two transition matrices were defined: the first one allowed us to estimate local survival probabilities, independently from the previous recording event, and the second allowed us to estimate recording probabilities, as a function of the previous recording event for each individual. We were not able to estimate separately the probability of using the pigeon house when alive probability of being recorded when using the pigeon house. The event matrix $E$ was therefore set to constant.

We conducted model selection to investigate the factors explaining local survival. We tested for the effect of age (for juvenile), time (year and month), colour score, and trap-dependence effect on recording probability. All interactions (up to four-way interactions) were tested alongside the corresponding main effects. Model selection was conducted according to the procedure used in Lebreton et al. (1992) using E-SURGE 1.7.1 (Choquet et al. 2009b). Local survival and recording probabilities were estimated using the maximum likelihood criterion. We first modelled recording probabilities ($p$), keeping the complete model for local survival probabilities ($\Phi$). Once the best model was obtained for $p$, we adjusted the model for $\Phi$. Models were compared using Akaike information criterion (AIC, Akaike 1981): models with lower AIC values are better supported by the data. When the difference between two AIC was less than two, the model with fewer parameters was preferred. AIC weights were calculated as they express the likelihood of the models relative to all the alternative models (Burnham and Anderson 2002). The beta parameters, and their significance levels following a Gaussian distribution, were estimated for each linear covariate and each level of categorical variables (Choquet et al. 2009b).

# Results

*Colour scores repartition*

The proportion of darker individuals was higher in adults than in juveniles in both pigeon houses (Pearson Chi-squared test; Fontenay: $\chi^2$ = 17.2, p = 0.0002; Pantin: $\chi^2$ = 5.92, p = 0.05): 1.5% of the juveniles in Fontenay and 8.5% in Pantin belonged to the darker morphs "T-pattern" and "Spread", compared to respectively 16.1% and 15.5% of the adults. There was a decrease over age in the proportion of intermediate "Checker" individuals in both pigeon houses, but no consistent trend for the paler morph "Blue Bar" (Table 2). The distribution of colour groups was similar between houses in adults ($\chi^2$ = 1.49, p = 0.47) but not in juveniles: more dark ("T-pattern" and "Spread") and less pale ("Blue Bar") juveniles were observed in Pantin than in Fontenay (Table 2, Pearson Chi-squared test with p-values computed by Monte Carlo simulations: $\chi^2$ = 7.78, p = 0.022).

*Local survival of adults*

Estimates of monthly adult local survival were similar between Fontenay (mean = 0.984; range = 0.937 – 1) and Pantin (mean = 0.990; range = 0.937 – 1). They only depended on month (in Pantin and Fontenay) and on year (in Fontenay), with no noticeable trend. The monthly or yearly number of chicks did not affect these parameters (Tables 3 and 6). In addition, adult local survival was not explained by plumage colouration (Tables 3 and 6).

*Local survival of juveniles in Fontenay*

Juvenile local survival in Fontenay differed between months and increased with age until five months old (best model with five age-classes; Table 4). It was also negatively correlated with the yearly number of chicks (Tables 4 and 6, Fig. 1). Moreover, juvenile local survival in

Fontenay was linked to colouration, with pale juveniles ("Blue Bar") having a lower local survival than darker juveniles ("Checker", "T-pattern" and "Spread"; Tables 4 and 6, Fig. 1).

*Local survival of juveniles in Pantin*

In Pantin, juvenile local survival was month-dependent and increased with age until three months old (best model with three age-classes; Table 4). However, there was no difference between colour morphs (Tables 4 and 6). We did not test for the effect of yearly number of chicks, since we only had data for two years.

## Discussion

The aim of the study was to compare local survival between differently coloured free-living feral pigeons. An increase in the proportion of darker individuals between juveniles and adults was noticed in both pigeon houses. Such differences have already been observed in Vienna (Haag-Wackernagel et al. 2006) and in different populations in and around Paris (Jacquin et al. 2013b), using a similar visual scoring protocol. These observations suggest directional selection at the juvenile stage, but could be flawed if detectability varies according to colouration (Gimenez et al. 2008).

Therefore, we estimated local survival in two pigeon houses using a mark-recapture protocol. Contrary to what we predicted, colouration was not related to local survival in adults. However, we found a relationship for juveniles: pale eumelanic juveniles had a lower monthly local survival probability than darker eumelanic ones. This relationship however was restricted to only one pigeon house (Fontenay). Thus, darker eumelanic juveniles could have a selective advantage compared to paler juveniles in some environmental contexts, which would explain the higher proportion of dark adult individuals in urban populations of feral pigeons compared to more rural ones. Interestingly, a long term study of barn owls *Tyto alba* also found a higher local survival of dark individuals in juveniles but not in adults (Roulin et al. 2010). Although other mechanisms (e.g. annual reproductive success) could generate fitness differences later in life, it thus seems that differences of survival and/or dispersal according to colouration are mainly expressed at the juvenile stage.

Adult local survival was similar between the two studied populations and was not linked to the number of chicks born. In contrast, local survival of juveniles was negatively correlated to the yearly number of chicks born in Fontenay, which varied according to egg removal management. Despite not being able to test this effect for juveniles in Pantin (due to having only two years of data), the observed decrease in local survival between 2008 and 2009 (Table 5) still coincided with an increase in the number of chicks. The observed variation in juvenile local survival between years may thus have resulted directly from variation in reproductive output due to chicks removal rather than from variation in environmental conditions. This contrasted response between adult and juveniles is consistent with previous studies that found juveniles to be more sensitive to environmental variation, in terms of survival (Gaillard et al. 2000) or dispersal (Schreiber et al. 2004). Taken together with the difference in colouration-dependence between adults and juveniles, this suggest that juvenile and adult local survival vary quite independently in feral pigeons, similarly to what was observed for other bird species (Altwegg et al. 2007).

The human-driven variation in chick number is likely to have induced changes in competition for food, which may impact juvenile mortality. Indeed, juveniles are less efficient foragers than adults, which makes them more sensitive to competition for food (Sol et al. 1998). A possible scenario is that when egg removal rate is high (i.e. before 2009 in Pantin, in 2009 in Fontenay), the number of chicks is lower, and juveniles are likely to experience a lower competition with other juveniles and/or with adults feeding them. Reduced competition could then act positively on local survival through lower mortality. Such effect could also be driven by differential natal dispersal rate. Indeed, juveniles tend to disperse towards less dense populations when densities on their natal site are high (Hetmański 2007).

Several underlying mechanisms could explain the difference in survival and/or dispersal between differently coloured juveniles. Indeed, melanin-based colouration intensity has been shown to be associated with various life-history traits such as behaviour, reproduction and parasite resistance in pigeons (Johnson and Johnston 1989, Johnston and Janiga 1995, Jacquin et al. 2011, 2012) and in other species (Moreno and Møller 2006, Ducrest et al. 2008, McKinnon and Pierrotti 2010). Different factors could thus act advantageously for dark coloured juvenile pigeons in one of our population. For instance, differential selection on pale and dark juveniles could be due to differences in competitive abilities. Darker individuals are known to be more aggressive and to have better competitive abilities for access to food resources (Senar 2006). Parasitism could also play a role in linking survival and colouration (Piault et al. 2009), because dark feral pigeons are known to develop stronger immune responses, to be more resistant to blood parasites (Jacquin et al. 2011) and to transmit more antibodies to their eggs (Jacquin et al. 2013a). As parasite transmission is generally higher in denser host populations (Arneberg 2001), darker eumelanic juveniles could have a selective advantage in dense populations. Selection could thus act directly through a higher survival and/or a lower dispersal of darker juveniles in dense populations. In agreement with this hypothesis, juvenile local survival differed according to colouration in Fontenay but not in Pantin, and competition is likely to be higher in Fontenay compared to Pantin due to lower reproductive control and a higher number of chicks born (Table 1). Moreover, differences in local survival between pale and dark juveniles in Fontenay were more pronounced in 2007 and 2008, when the number of chicks born was high, compared to 2009, where the number of chicks was lower (Table 3, Fig 1). An alternative explanation could be the higher urbanisation rate in Fontenay compared to Pantin, resp. 92.4% and 85.2% (Jacquin et al. 2013b). However the effect of urbanisation could only be tested by studying several sites in a wide range of urbanisation rates. In birds, colouration-by-environment interactions have already been found on chicks' condition and growth, taken as proxies for survival prospect (Roulin et al. 2008, Piault et al. 2009). However a colouration-by-environment interaction on dispersal has only been described in insects (Ahnesjö and Forsman 2006).

Our results suggest that eumelanin-based colouration is under directional survival selection in dense populations of highly urbanized areas, and that such selection occurs through differential survival or dispersal rate of juveniles. Interactions between melanin-based colouration and population densities could explain the observed differences in local survival. Consistently, Obukhova (2011) found that proportions of darker morphs were positively correlated with pigeon densities both in space and time. Darker eumelanic colouration could thus reflect an adaptation to high population density, which could entail a selective advantage in more urbanized areas. Indeed, densely urbanized areas are associated with higher densities of pigeon populations compared to more rural ones (Johnston and Janiga 1995), and stronger competition for food and changes in parasite exposure has often been invoked as a central selective factor in such densely colonized urban areas (Shochat et al. 2004, Bradley and Altizer 2004).

To conclude, this study provides the first demonstration of colouration-dependent selection in juvenile feral pigeons taking into account detection probabilities. It contributes to the growing body of evidence that darker eumelanic individuals have a higher survival and/or a lower dispersal than paler ones, especially at the juvenile stage and in stressful environments (Roulin et al. 2008, Piault et al. 2009). The observed differences in local survival according to colouration intensity were indeed context-dependent, linked potentially to the degree of competition and/or parasite exposure in more or less dense populations. Although our results are consistent with such density-dependent effects, the present study does not provide enough data to test them in full. Moreover, it was conducted in populations exploiting pigeon houses, so conditions might be quite different from those experienced by free-living populations. A current survey of colour-ringed pigeons from free-living populations and pigeon houses in various sites of Paris area may yield interesting information on the environmental correlates of colouration-dependent local survival and on the relative contributions of mortality and dispersal.

*Acknowledgments* – All experiments and protocols were approved and authorized by the veterinary department of Seine-et-Marne (authorization N°77-05). We are grateful to the AERHO association, and especially to Catherine Dehay and Pascale Beauvois, for conducting the captures and collecting data, and to Maryline Berthaud and Romain Lorrillière for computing the data before analysis. We would like to thank three anonymous reviewers for their comments on a previous version of this manuscript. This work was supported by grants from the Region Ile-de-France (Sustainable Development Network R2DS N° 2008-07). C. Récapet was supported by grants from the French Ministry of Research. L. Jacquin was supported by the French Ministry of Research and the Fyssen foundation. L. Dauphin was supported by an ATM program in the French National Museum of Natural History.

# References

Ahnesjö, J. and Forsman, A. 2006. Differential habitat selection by Pygmy Grasshopper colour morphs, interactive effects of temperature and predator avoidance. – Evol. Ecol. 20: 235–257.

Akaike, H. 1981. Likelihood of a model and information criteria. – J. Econom. 16: 3–14.

Almasi, B., Jenni, L., Jenni-Eiermann, S. and Roulin, A. 2010. Regulation of stress response is heritable and functionally linked to melanin-based coloration. – J. Evol. Biol. 23: 987–996.

Altwegg, R., Schaub, M. and Roulin, A. 2007. Age-specific fitness components and their temporal variation in the Barn Owl. – Am. Nat. 169: 47–61.

Arneberg, P. 2001. An ecological law and its macroecological consequences as revealed by studies of relationships between host densities and parasite prevalence. – Ecography, 24: 352–358.

Berry, B. J. L. 1990. Urbanization. – In: Turner B. L. (ed). The Earth as transformed by human action. Cambridge University Press, pp. 103–119.

Blair, R. 2004. The Effects of urban sprawl on birds at multiple levels of biological organization. – Ecol. Soc. 9: 2.

Bradley, C. A. and Altizer S. 2007. Urbanization and the ecology of wildlife diseases. – Trends Ecol. Evol. 22: 95–102.

Brownie, C. and Robson, D. S. 1983. Estimation of time-specific survival rates from tag-resighting samples: a generalization of the Jolly-Seber model. – Biometrics 39: 437–453.

Burnham, K. P. and Anderson, D. R. 2002. Model selection and multimodel inference: a practical information-theoretic approach. Springer, New York.

Choquet, R., Lebreton, J.-D., Gimenez, O., Reboulet, A.-M. and Pradel, R. 2009. U-CARE: Utilities for performing goodness of fit tests and manipulating CApture-REcapture data. – Ecography 32: 1071–1074.

Choquet, R., Rouan, L. and Pradel, R. 2009. Program E-SURGE: A Software application for fitting multievent models. – In: Thomson, D. L., Cooch, E. G. and Conroy, M. J. (eds). Environmental and Ecological Statistics Vol.3. Springer US, pp. 845–865.

Clergeau, P., Savard, J. P. L., Mennechez, G. and Falardeau, G. 1998. Bird abundance and diversity along an urban-rural gradient: a comparative study between two cities on different continents. – Condor 100: 413–425.

Ducrest, A.-L., Keller, L. and Roulin, A. 2008. Pleiotropy in the melanocortin system, colouration and behavioural syndromes. – Trends Ecol. Evol. 23: 502–510.

Erz, W. 1964. Populationsökologische Untersuchungen an der Avifauna zweier norddeutscher Großstädte. – Z. Wiss. Zool. 170: 1–111.

Evans, K. L., Gaston, K. J., Sharp, S. P., McGowan, A., Simeoni, M. and Hatchwell, B. J. 2009. Effects of urbanisation on disease prevalence and age structure in Blackbird *Turdus merula* populations. – Oikos 118: 774–782.

Fokidis, H. B., Greiner, E. C. and Deviche, P. 2008. Interspecific variation in avian blood parasites and haematology associated with urbanization in a desert habitat. – J. Avian Biol. 39: 300–310.

French, S. S., Fokidis, H. B. and Moore, M. C. 2008. Variation in stress and innate immunity in the Tree Lizard (*Urosaurus ornatus*) across an urban-rural gradient. J. Comp. Physiol. B – 178: 997–1005.

Gaillard, J.-M., Festa-Bianchet, M., Yoccoz, N. G., Loison, A. and Toïgo, C. 2000. Temporal variation in fitness components and population dynamics of large herbivores. – Annu. Rev. Ecol. Syst. 31: 367–93.

Germaine, S. S., Rosenstock, S. S., Schweinsburg, R. E., and Richardson, W. S. 1998. Relationships among breeding birds, habitat and residential development in greater Tucson, Arizona. – Ecol. Appl. 8: 680–691.

Gimenez, O., Viallefont, A., Charmantier, A., Pradel, R., Cam, E., Brown, C. R., Anderson, M. D., Bomberger Brown, M., Covas, R., and Gaillard, J.-M. 2008. The Risk of flawed inference in evolutionary studies when detectability is less than one. – Am. Nat. 172: 441–448.

Greenwood, P. J. and Harvey, P. H. 1982. The natal and breeding dispersal of birds. – Annu. Rev. Ecol. Syst. 13: 1–21.

Grimm, N. B., Grove, J. M., Pickett, S. T. A. and Redman, C. L. 2000. Integrated approaches to long-term studies of urban ecological systems. – BioScience 50: 571–584.

Haag-Wackernagel, D. 1998. Die Taube. Vom heiligen Vogel der Liebesgöttin zur Strassentaube. Schwabe and Co.

Haag-Wackernagel, D., Heeb, P. and Leiss, A. 2006. Phenotype-dependent selection of juvenile urban feral Pigeons *Columba livia*. – Bird Study 53: 163–170.

Haase, E., Ito, S., Sell, A. and Wakamatsu, K. 1992. Melanin concentrations in feathers from wild and domestic Pigeons. – J. Hered. 83: 64–67.

Hetmański, T. 2007. Dispersion asymmetry within a feral Pigeon *Columba livia* population. – Acta Ornithologica 42: 23–31.

Jacquin, L., Cazelles, B., Prévot-Julliard, A.-C., Leboucher, G. and Gasparini, J. 2010. Reproduction management affects breeding ecology and reproduction costs in feral urban pigeons (*Columba livia*). – Can. J. Zool. 88: 781–787.

Jacquin, L. 2011. Coloration mélanique et stratégies d'histoire de vie chez le pigeon biset urbain. – PhD thesis, Université Pierre et Marie Curie, Paris.

Jacquin, L., Lenouvel, P., Haussy, C., Ducatez, S. and Gasparini, J. 2011. Melanin-based colouration is related to parasite intensity and cellular immune response in an urban free-living bird: the feral Pigeon *Columba livia*. – J. Avian Biol. 42: 11–15.

Jacquin, L., Récapet, C., Bouche, P., Leboucher, G. and Gasparini, J. 2012. Melanin-based colouration reflects alternative strategies to cope with food limitation in Pigeons. – Behav. Ecol. 23: 907–915.

Jacquin, L., Haussy, C., Bertin, C., Laroucau, K., and Gasparini, J. 2013a. Darker female pigeons transmit more specific antibodies to their eggs than do paler ones. – Biol. J. Linn. Soc. 108: 647–657.

Jacquin, L., Récapet, C., Prevot-Julliard, A.-C., Leboucher, G., Lenouvel, P., Erin, N., Frantz, A., Corbel, H., and Gasparini, J. 2013b. A potential role for parasites in the maintenance of bird color polymorphism in cities. – Oecologia. Advance online publication. DOI 10.1007/s00442-013-2663-2.

Johnston, R. F. and Janiga, M. 1995. Feral Pigeons. Oxford University Press.

Johnson, S. G. and Johnston, R. F. 1989. A multifactorial study of variation in interclutch interval and annual reproductive success in the feral Pigeon, *Columba livia*. – Oecologia 80: 87–92.

Kark, S., Iwaniuk, A., Schalimtzek, A. and Banker, E. 2007. Living in the city: Can anyone become an 'urban exploiter'? – J. Biogeogr. 34: 638–651.

Lebreton, J.-D., Burnham, K. P., Clobert, J. and Anderson, D. R. 1992. Modeling survival and testing biological hypotheses using marked animals: a unified approach with case studies. – Ecol. Monogr. 62: 67–118.

Leiss, A. and Haag-Wackernagel, D. 1999. Gefiederfärbungen bei der Stragentaube (*Columba livia*). – J. Ornithol. 140: 341–353.

Luniak, M. 2004. Synurbization - Adaptation of animal wildlife to urban development. – In: Shaw et al. (ed). Proceedings 4th International Urban Wildlife Symposium, pp. 50–55.

McKinnon, J. S. and Pierotti, M. E. R. 2010. Colour polymorphism and correlated characters: Genetic mechanisms and evolution. – Mol. Ecol. 19: 5101–5125.

Moreno, J. and Møller, A. P. 2006. Are melanin ornaments signals of anti-oxidant and immune capacity in birds? – Acta Zoologica Sinica 52: 202–208.

Møller, A. P. 2009. Successful city dwellers: a comparative study of the ecological characteristics of urban birds in the western Palearctic. – Oecologia 159: 849–858.

Møller, A. P. 2010. Interspecific variation in fear responses predicts urbanization in birds. – Behav. Ecol. 21: 365–371.

Obukhova, N. Y. 2001. Geographic variation of colour in the synanthropic Blue Rock Pigeon. – Russ. J. Genet. 37: 649–658.

Obukhova, N. Y. 2007. Polymorphism and phene geography of the Blue Rock Pigeon in Europe. – Russ. J. Genet. 43: 492–501.

Obukhova, N. Y. 2011. Dynamics of balanced polymorphism morphs in Blue Rock Pigeon *Columbia livia*. – Russ. J. Genet. 47: 83–89.

Partecke, J., Van't Hof, T. and Gwinner, E. 2004. Differences in the timing of reproduction between urban and forest European Blackbirds (*Turdus merula*): Result of phenotypic flexibility or genetic differences? – Proc. Biol. Sci. 271: 1995–2001.

Piault, R., Gasparini, J., Roulin, A., Bize, P., Jenni-Eiermann, S., and Roulin, A. 2009. Pheomelanin-based coloration and the ability to cope with variation in food supply and parasitism. – Am. Nat. 174: 548–556.

Pickett, S. T. A., Burch, W. R., Dalton, S. E., Foresman, T. W., Grove, J. M. and Rowntree, R. 1997. A Conceptual framework for the study of human ecosystems in urban areas. – Urban Ecosyst. 1: 185–199.

Pollock, K. H., Hines, J. E. and Nichols, J. D. 1985. Goodness–of–fit tests for open capture–recapture models. – Biometrics 41: 399–410.

Pradel, R. 1993. Flexibility in survival analysis from recapture data: Handling trap-dependence. – In: Lebreton, J.D. and North, P. M. (eds). Marked individuals in the study of bird population. Birkhaeuser-Verlag , pp. 9–28.

Quesada, J., and Senar, J.C. 2007. The role of melanin- and carotenoid-based plumage coloration in nest defence in the great tit. – Ethology 113: 640–647.

Roulin, A., Altwegg, R., Jensen, H., Steinsland, I., and Schaub, M. 2010. Sex-dependent selection on an autosomal melanic female ornament promotes the evolution of sex ratio bias. – Ecol. Lett. 13: 616–626.

Roulin, A., Gasparini, J., Bize, P., Ritschard, M., and Richner, H. 2008. Melanin-based colorations signal strategies to cope with poor and rich environments. – Behav. Ecol. Soc. 62: 507–519.

Roulin, A., and Altwegg, R. 2007. Breeding rate is associated with pheomelanism in male and with eumelanism in female barn owls. – Behav. Ecol. 18: 563–570.

Roulin, A. 2004. The Evolution, maintenance and adaptive function of genetic colour polymorphism in birds. – Biol. Rev. Camb. Philos. Soc. 79: 815–848.

Roulin, A., Jungi, T. W., Pfister, H., and Dijkstra, C. 2000. Female Barn owls (*Tyto alba*) advertise good genes. – Proc. R. Soc. B 267 1446: 937–941.

Reyer, H. U., Fischer, W., Steck, P., Nabulon, T., and Kessler, P. 1998. Sex-specific nest defense in house sparrows (Passer domesticus) varies with badge size of males. – Behav. Ecol. Soc. 42: 93–99.

Saino, N., Romano, M., Rubolini, D., Ambrosini, R., Caprioli, M., Milzani, A., Costanzo, A., Colombo, G., Canova, L. and Wakamatsu, K. 2013. Viability is associated with melanin-based coloration in the Barn Swallow (*Hirundo rustica*). – PLoS ONE 8: e60426.

Schreiber, E. A., Doherty, P. F. Jr. and Schenk, G. A. 2004. Dispersal and survival rates of adult and juvenile Red–tailed tropicbirds (Phaethon rubricauda) exposed to potential contaminants. – Anim. Biodiv. Cons. 27: 531–540.

Senar, J. C. 2006. Bird colours as intrasexual signals of aggression and dominance. In: Hill, G. E. and McGraw, K. J. (eds). Bird colouration II. Function and evolution. Harvard University Press, pp. 87–136.

Shochat, E. 2004. Credit or debit? Resource input changes population dynamics of city-slicker birds. – Oikos 106: 622–626.

Shochat, E., Warren, P. S., Faeth, S. H., McIntyre, N. E. and Hope, D. 2006. From Patterns to emerging processes in mechanistic urban ecology. – Trends Ecol. Evol. 21: 186–191.

Sol, D., Santos, D. M., Garcia, J. and Cuadrado, M. 1998. Competition for food in urban Pigeons: The Cost of being juvenile. – Condor 100: 298–304.

van den Brink, V., Dreiss, A. N., and Roulin, A. 2012. Melanin-based coloration predicts natal dispersal in the barn owl, *Tyto alba.* – Anim. Behav. 84: 805–812.

Yeh, P. J. 2004. Rapid evolution of a sexually selected trait following population establishment in a novel area. – Evolution 58: 166–174.

1  **Table 1:** Description of the two pigeon houses surveyed: egg-removal protocol (0, 1 or 2 eggs removed per clutch), total number of chicks born

2  per year, and months during which recording took place (in grey).

|  | Year | 2007 | 2008 | 2009 |
|---|---|---|---|---|
|  | Month | J F M A M J J A S O N D | J F M A M J J A S O N D | J F M A M J J A S O N D |
| Pantin | Egg removal | 2 eggs per clutch | 1 egg per clutch | 1 egg per clutch |
| Pantin | Chicks born | 1 | 61 | 86 |
| Pantin | Recording | | | |
| Fontenay | Egg removal | 0 egg per clutch | 0 egg per clutch | 1 egg per clutch |
| Fontenay | Chicks born | 98 | 104 | 34 |
| Fontenay | Recording | | | |

4  **Table 2:** Number of birds included in the survey. All were recorded at least once after ringing. The percentage of each colour morph relative to
5  the number of individuals of the same age and site are given in brackets.

|  |  |  | Colour morph |  |  |
|---|---|---|---|---|---|
|  |  | All | Blue Bar | Checker | T-pattern/Spread |
| Adults | Pantin | 188 | 51 (27.1) | 107 (56.9) | 30 (16.0) |
|  | Fontenay | 136 | 29 (21.3) | 85 (62.5) | 22 (16.2) |
| Juveniles | Pantin | 89 | 17 (19.1) | 64 (71.9) | 8 (9.0) |
|  | Fontenay | 128 | 35 (27.3) | 91 (71.1) | 2 (1.6) |

6  **Table 3**: Model selection for adult local survival (selected models in bold). All models with an AIC weight higher than 1% are reported. In
7  Pantin, recapture probabilities are set as year-dependent and include trap-dependence. In Fontenay, recapture probabilities are set as fully time-
8  dependent (month x year) and include trap-dependence. Survival probabilities all include a transience-effect (see methods). Models with
9  interactions include the corresponding main effects.

| Site | Modelled effects on adult local survival | Parameters | Deviance | AIC | AIC weight |
|---|---|---|---|---|---|
| *Pantin* | Yearly number of chicks + Month | 19 | 701.9 | 739.9 | 0.323 |
| | **Month** | **18** | **704.3** | **740.3** | **0.265** |
| | Year + Month | 20 | 700.3 | 740.3 | 0.256 |
| | Colour + Yearly number of chicks + Month | 21 | 701.7 | 743.7 | 0.048 |
| | Year × Month | 32 | 680.2 | 744.2 | 0.037 |
| | Colour × Yearly number of chicks + Month | 23 | 699.3 | 745.3 | 0.021 |
| | Colour + Year + Month | 22 | 701.6 | 745.6 | 0.018 |
| | Yearly number of chicks | 8 | 730.3 | 746.3 | 0.013 |
| *Fontenay* | **Year × Month** | **59** | **754.1** | **872.1** | **0.738** |
| | Colour + Year × Month | 61 | 753.2 | 875.2 | 0.152 |
| | Month | 44 | 789.8 | 877.8 | 0.041 |
| | Year + Month | 46 | 786.5 | 878.5 | 0.029 |
| | Yearly number of chicks + Month | 45 | 789.2 | 879.2 | 0.020 |

11  **Table 4**: Model selection for juvenile local survival (selected models in bold). For each model, only the model with the best age structure is
12  reported. Only models with an AIC weight higher than 2% are reported. In Pantin, recapture probability is modelled as age-dependent up to 4
13  months. In Fontenay, recapture probability is dependent on year and age (up to 4 months), with trap-dependence. Models with interactions
14  include the corresponding main effects.

| Site | Modelled effects on juvenile local survival | Parameters | Deviance | AIC | AIC weight |
|---|---|---|---|---|---|
| *Pantin* | **Month + Age up to 3 months** | **20** | **291.0** | **331.0** | **0.095** |
| | Year + Month + Age up to 3 months | 21 | 289.9 | 331.9 | 0.061 |
| | Colour + Month + Age up to 3 months | 21 | 290.3 | 332.3 | 0.051 |
| | Colour × Age up to 3 months + Year + Month | 24 | 284.1 | 332.1 | 0.055 |
| | Colour + Year + Month + Age up to 3 months | 22 | 289.3 | 333.3 | 0.031 |
| | (Year + Age up to 3 months) × Colour + Month | 25 | 283.5 | 333.5 | 0.027 |
| *Fontenay* | Colour + Year + Month + Age up to 5 months | 28 | 648.5 | 704.5 | 0.256 |
| | **Colour + Yearly number of chicks + Month + Age up to 5 months** | **27** | **651.3** | **705.3** | **0.166** |
| | Colour + Yearly number of chicks × Age up to 5 months + Month | 27 | 653.9 | 707.9 | 0.045 |
| | Colour + Yearly number of chicks + Age up to 5 months | 15 | 678.0 | 708.0 | 0.043 |
| | Colour + Month + Age up to 5 months | 26 | 656.2 | 708.2 | 0.040 |
| | Colour × Age up to 6 months + Year + Month | 34 | 640.3 | 708.3 | 0.038 |
| | Colour × Age up to 6 months + Yearly number of chicks + Month | 33 | 643.5 | 709.5 | 0.021 |

16  **Table 5:** Maximum-likelihood estimates and 95% confidence intervals for monthly local survival rates according to year, site and age. Estimates
17  for both colour morphs are given separately for juveniles of Fontenay ("Pale": Blue Bar, "Dark": Checker/T-pattern/Spread). These values are
18  obtained from the best models without the month effect.

|  |  |  | Year | | |
|---|---|---|---|---|---|
|  |  |  | 2007 | 2008 | 2009 |
| Adults | Pantin |  | 0.99 [0.46 – 1.00] | 0.99 [0.26 – 1.00] | 0.99 [0.86 – 0.99] |
|  | Fontenay |  | 0.98 [0.96 – 0.99] | 0.98 [0.96 – 0.98] | 0.99 [0.97 – 0.98] |
| Juveniles | Pantin |  | – | 0.90 [0.78 – 0.96] | 0.59 [0.49 – 0.68] |
|  | Fontenay | Pale | 0.55 [0.41 – 0.69] | 0.48 [0.31 – 0.66] | 0.85 [0.70 – 0.93] |
|  |  | Dark | 0.78 [0.70 – 0.85] | 0.73 [0.59 – 0.84] | 0.94 [0.87 – 0.97] |

**Table 6:** Maximum-likelihood estimates ± S.E. (beta parameters) for the effects of colouration and of the yearly number of chicks born on local survival. P-values are given into brackets.

|  |  | Colouration (relative to Blue Bar birds) | | Number of chicks born |
|---|---|---|---|---|
| Adults |  | Checker | T-pattern/Spread |  |
|  | Pantin | + 0.15 ± 0.50 (0.76) | + 0.30 ± 0.73 (0.68) | - 3.74 ± 2.70 (0.17) |
|  | Fontenay | - 0.05 ± 0.38 (0.90) | + 0.31 ± 0.52 (0.54) | + 0.80 ± 1.01 (0.43) |
| Juveniles |  | Checker/T-pattern/Spread | | |
|  | Pantin | + 0.37 ± 0.42 (0.39) | | – |
|  | Fontenay | **+ 1.27 ± 0.34 (0.0002)** | | - 3.32 ± 1.77 (0.06) |

25   **Figure 1**: Monthly mortality rate of one-month old juveniles in Fontenay as a function of year and colouration ("Pale": Blue Bar, "Dark":
26   Checker/T-pattern/Spread) and number of chicks born in Fontenay (dotted line) as a function of year. Model selection indicated that mortality
27   differed between years, as a function of the yearly number of chicks born, as well as between colour scores. The graphs for older juveniles
28   indicate similar trends. The error bars give the 95% confidence intervals.
29
30

31

32    Fig.1

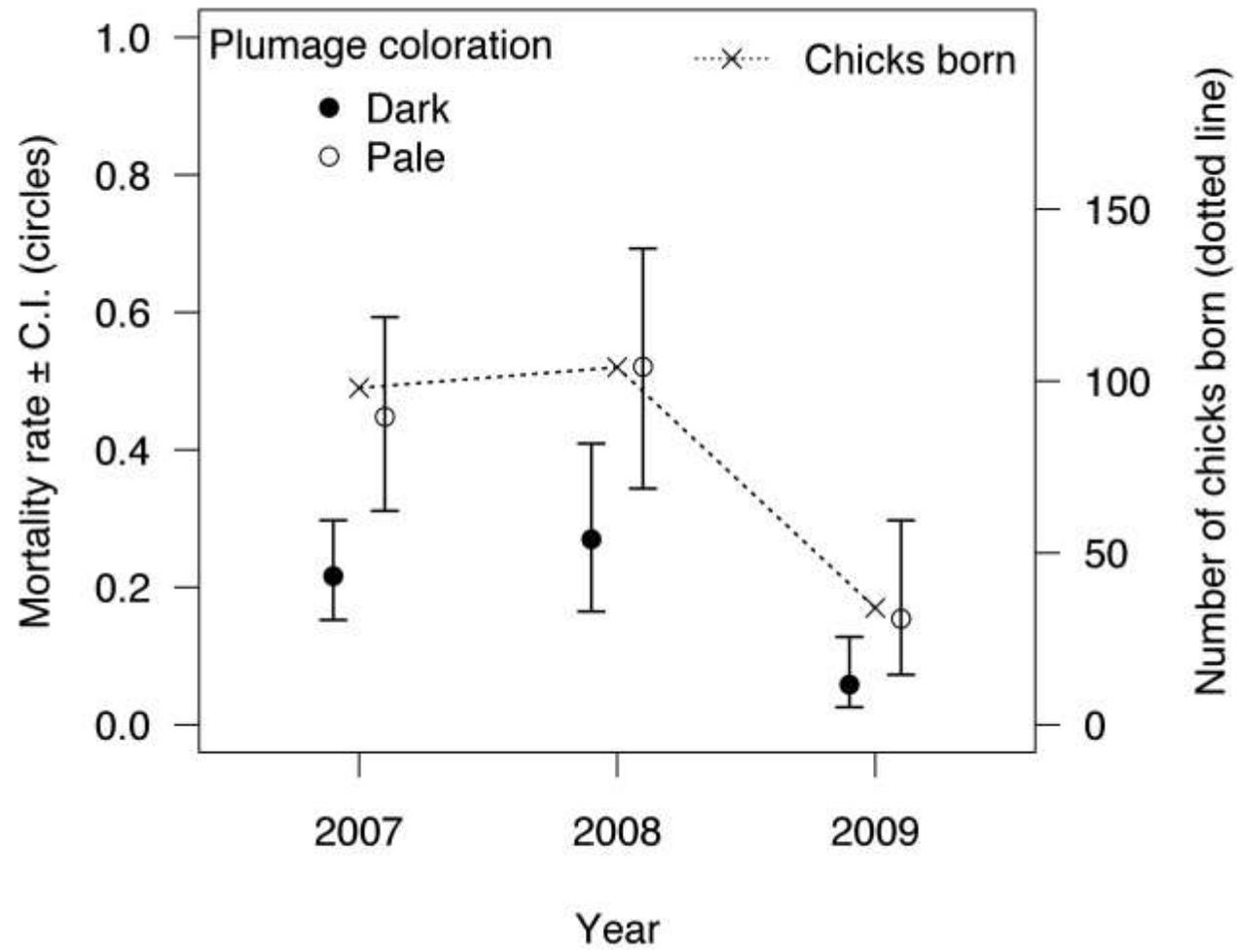

33